\documentclass[aps,prb,twocolumn,floats,superscriptaddress,preprintnumbers,amsmath,amssymb]{revtex4}

\usepackage{pslatex}
\usepackage{graphicx} 
\usepackage{dcolumn} 
\usepackage{bm} 

\begin{document}

\preprint{(submitted to PCCP)}

\title{Azobenzene versus 3,3',5,5'-tetra-\emph{tert}-butyl-azobenzene (TBA) at Au(111):\\
Characterizing the role of spacer groups}

\author{Erik R. McNellis}
\affiliation{Fritz-Haber-Institut der Max-Planck-Gesellschaft, Faradayweg 4-6, D-14195 Berlin (Germany)}

\author{Christopher Bronner}
\affiliation{Fachbereich Physik, Freie Universit{\"a}t Berlin,
Arnimallee 14, D-14195 Berlin, (Germany)}

\author{J\"org Meyer}
\affiliation{Fritz-Haber-Institut der Max-Planck-Gesellschaft, Faradayweg 4-6, D-14195 Berlin (Germany)}

\author{Martin Weinelt}
\affiliation{Fachbereich Physik, Freie Universit{\"a}t Berlin,
Arnimallee 14, D-14195 Berlin, (Germany)}
\affiliation{Max-Born-Institut, Max-Born-Str. 2A, D-12489 Berlin (Germany)}

\author{Petra Tegeder}
\affiliation{Fachbereich Physik, Freie Universit{\"a}t Berlin,
Arnimallee 14, D-14195 Berlin, (Germany)}

\author{Karsten Reuter}
\affiliation{Fritz-Haber-Institut der Max-Planck-Gesellschaft, Faradayweg 4-6, D-14195 Berlin (Germany)}
\affiliation{Department Chemie, Technische Universit{\"a}t M{\"u}nchen, Lichtenbergstr. 4, D-85747 Garching (Germany)}


\begin{abstract}
We present large-scale density-functional theory (DFT) calculations and temperature programmed desorption measurements to characterize the structural, energetic and vibrational properties of the functionalized molecular switch $3,3',5,5'$-tetra-\emph{tert}-butyl-azobenzene (TBA) adsorbed at Au(111). Particular emphasis is placed on exploring the accuracy of the semi-empirical dispersion correction approach to semi-local DFT (DFT-D) in accounting for the substantial van der Waals component in the surface chemical bond. In line with previous findings for benzene and pure azobenzene at coinage metal surfaces, DFT-D significantly overbinds the molecule, but seems to yield an accurate adsorption geometry as far as can be judged from the experimental data. Comparing the {\em trans} adsorption geometry of TBA and azobenzene at Au(111) reveals a remarkable insensitivity of the structural and vibrational properties of the $-\!{\rm N}\!\!=\!\!{\rm N}\!-$ moiety. This questions the established view of the role of the bulky tert-butyl-spacer groups for the switching of TBA in terms of a mere geometric decoupling of the photochemically active diazo-bridge from the gold substrate.
\end{abstract}

\maketitle

\section{Introduction}

The ultimate goal of nanotechnology is control over molecular-scale mechanical and electronic components. An oft-proposed route to this goal is the construction of components from controllable single molecules. An important class of such molecules with obvious applications is formed by those that have properties that are reversibly and bi-stably modifiable by external stimuli -- so-called \emph{molecular switches}. One example is the azobenzene molecule (H$_{5}$C$_{6}$-N$\!\!=\!\!$N-C$_{6}$H$_{5}$), which can be bi-stably photo-isomerized between its planar, $C_{2h}$ symmetric \emph{trans} and torsioned-twisted, $C_{2}$ \emph{cis} conformers in both solution and gas-phase. The high yield and stability of this reaction have rendered azobenzene an archetype of molecular switch research with proposed technical applications including e.g.~light-driven actuators\cite{yu03} and information storage media\cite{liu90,ikeda95}.

\begin{figure}
\begin{centering}
\includegraphics[width=1\columnwidth]{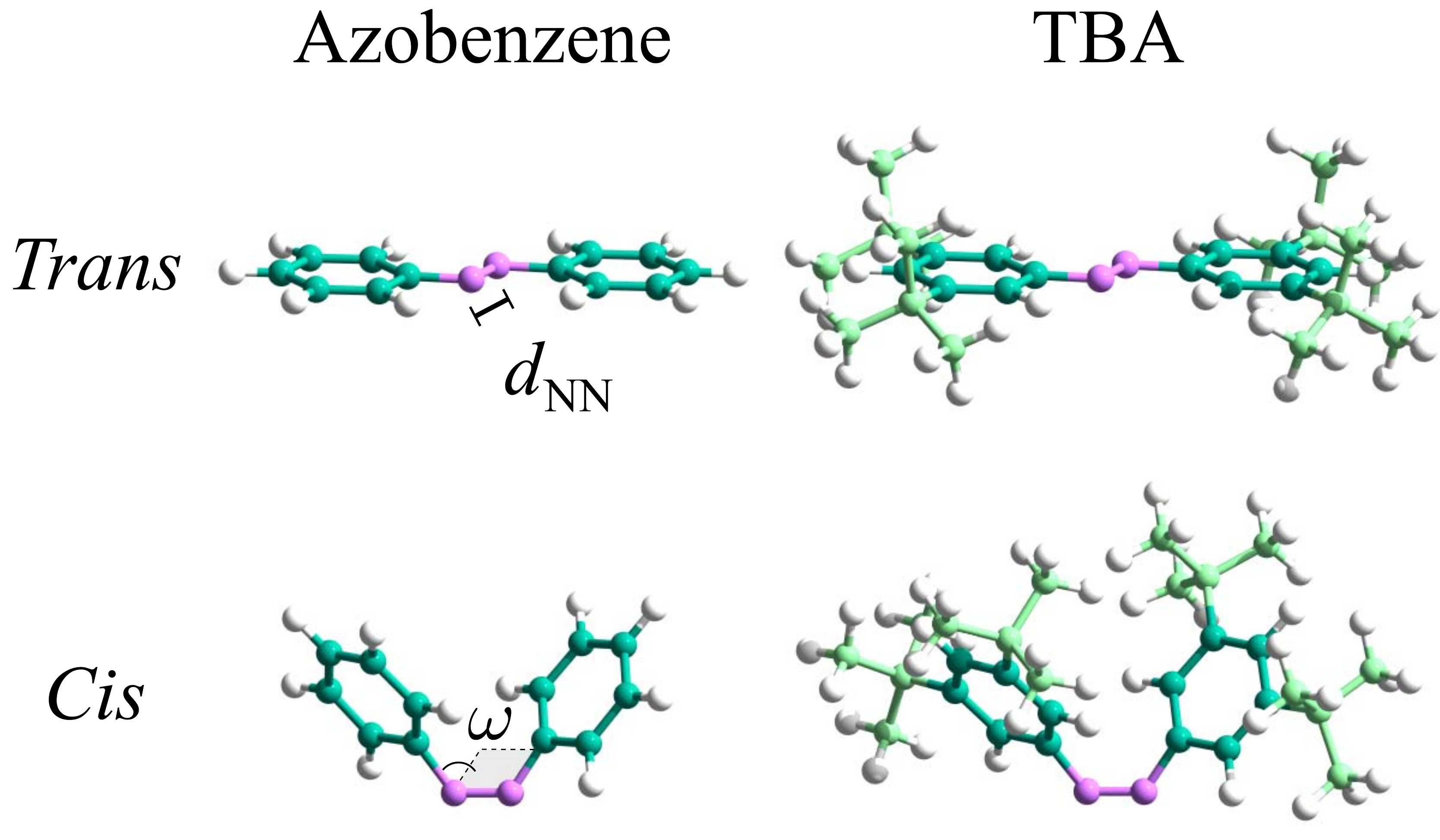}
\end{centering}
\caption{\label{fig1}
Perspective views of {\em trans} and {\em cis} azobenzene and TBA, together with an illustration of the diazo-bridge bond length $d_{\rm NN}$ and the dihedral CNNC angle $\omega$. The latter is defined as the smallest angle between two planes spanned by the $-\!{\rm N}\!\!=\!\!{\rm N}\!-$ bridge and $-$C$-$N$-$ and $-$N$-$C$-$ bonds to the first and second phenyl-ring, respectively. The C atoms of the phenyl-rings have been darkened to allow easier distinction of azobenzene backbone and functional butyl groups in TBA.}
\end{figure}

For many such applications, switching of molecules at solid interfaces -- adsorbed at metal surfaces, for example -- is of particular interest. Unfortunately however, the switching properties of azobenzene have proven highly sensitive to the adsorbate-substrate interaction: Even at nearly chemically inert close-packed noble metal surfaces, switching of surface-adsorbed azobenzene by light has never been achieved, and switching by excitation with a scanning tunneling microscope (STM) tip has been successful only at Au(111)\cite{choi06}. A natural route to restoring the adsorbate switching ability is to further decouple the frontier $\pi,n$ and $\pi^{*}$ orbitals responsible\cite{ishikawa01,cembran04} for the gas- or liquid phase photo-isomerization from the substrate electronic structure. With these frontier orbitals largely located at the central diazo ($-$N=N$-$) bridge an intuitive idea to achieve such a decoupling is to functionalize the molecule with bulky spacer groups that prevent a closer encounter of the photochemically active unit with the substrate. This is precisely the notion behind the arguably to date most studied such adsorbate, $3,3',5,5'$-tetra-\emph{tert}-butyl-azobenzene (TBA)
\cite{jung97,moresco01,alemani06}. TBA consists of azobenzene functionalized with four {\em tert}-butyl (-C-(CH$_{3}$)$_{3}$) groups in the phenyl-ring meta positions as illustrated in Fig. \ref{fig1}. These 'table legs' were indeed found to enhance the switching efficiency of the adsorbed species, as e.g. indicated by the successful TBA switching by light at Au(111) \cite{hagen07,comstock07}. However, attempts at STM-tip induced switching at ostensibly comparable substrates such as Ag(111)\cite{tegeder07} and Au(100)\cite{alemani08} have been unsuccessful, indicating that TBA is not significantly more robust to specifics of the substrate interaction than pure azobenzene.

These circumstances beg the question, how and to what degree the TBA butyl groups really 'decouple' the photochromic moiety from the substrate. For a corresponding atomic-scale understanding the detailed characterization of the adsorbate geometry and binding constitutes a prerequisite and is the main objective of the present contribution. In contemporary surface science, corresponding analyses are increasingly performed by quantitative first-principles electronic structure calculations. Particularly density-functional theory (DFT) with present-day local or semi-local exchange-correlation (xc) functionals has developed into an unparalleled workhorse for this task, with often surprising accuracy particularly with respect to structural properties of the surface adsorption system. For TBA at Au(111) corresponding calculations are already challenged by the extension of the functionalized molecule and the simultaneous necessity to describe the metal band structure within a periodic supercell approach. On a more fundamental level, the real limitation comes nevertheless from sizable dispersive van der Waals (vdW) contributions to the surface chemical bond as characteristic for organic molecules containing highly polarizable aromatic ring systems \cite{jenkins09}. With local and semi-local xc functionals inherently unable to account for such contributions \cite{kristyan94} skewed, if not qualitatively wrong results must therefore be suspected. We have recently quantified this for azobenzene at the close-packed coinage metal surfaces \cite{mcnellis09,mcnellis09_2}. Comparison to detailed structural and energetic reference data from normal-incidence x-ray standing wave (NI-XSW) and temperature programmed desorption (TPD) measurements for azobenzene at Ag(111) demonstrates that the prevalent semi-local DFT xc treatment leads indeed to a significant underbinding with key structural parameters deviating by more than 0.5\,{\AA} \cite{mercurio10}. 

For system sizes as those implied by the adsorption of azobenzene or TBA an appealing and computationally tractable possibility to improve on this situation is a semi-empirical account of dispersive interactions within the framework of so-called DFT-D schemes \cite{mcnellis09_2,mercurio10,wu02,grimme04,jurecka07,atodiresei08,tkatchenko09}. In this approach the vdW interactions not described by present-day xc functionals are approximately considered by adding a pairwise interatomic $C_6 R^{-6}$ term to the DFT energy. At distances below a cut-off, motivated by the vdW radii of the atom pair, this long-range dispersion contribution is heuristically reduced to zero by multiplication with a short-range damping function. While the applicability of this approach to adsorption at metal surfaces is uncertain ({\em vide infra}), our recent benchmark study for azobenzene at Ag(111) revealed that in particular the most recent DFT-D scheme due to Tkatchenko and Scheffler (TS) \cite{tkatchenko09} yields excellent structural properties, albeit at a notable overbinding \cite{mercurio10}. 

In this contribution we further explore the generality of this finding by analyzing the adsorption geometry, vibrations and energetics of TBA at Au(111). Comparison against our recent near edge x-ray absorption fine structure (NEXAFS) \cite{schmidt10} and high-resolution electron energy loss spectroscopy (HREELS) \cite{ovari07} measurements, as well as a complete analysis of new TPD data confirms the accurate structural and vibrational predictions reached by the DFT-D TS scheme. The again obtained significant overbinding furthermore supports the interpretation \cite{mercurio10} that the neglect of metallic screening is the main limitation in the application of this scheme to the adsorption of organic molecules at metal surfaces. Comparing the TBA data to those for pure azobenzene at Au(111) we find a qualitatively different adsorption geometry for the functional backbone in the case of the {\em cis} isomer. For the {\em trans} isomer, on the other hand, we determine an intriguing structural and vibrational insensitivity of the photochemically active central diazo-bridge to the presence of the bulky spacer groups. The role of the latter for the switching efficiency is therefore more subtle than simple geometric decoupling and will be the topic of continuing work in our groups.

\section{Method}

\subsection{Theory}

The DFT-D methodology followed in this work corresponds exactly to that employed in our preceding work for the adsorption of azobenzene. We therefore restrict ourselves here to a brief account and refer to our previous publications \cite{mcnellis09,mcnellis09_2,mercurio10} for an in-depth description of the underlying concepts and technical details.

The DFT calculations were performed using a plane-wave basis set and library ultrasoft pseudopotentials\cite{vanderbilt90} as implemented in and distributed with the CASTEP\cite{clark05} code. The generalized gradient approximation (GGA) to the xc-functional with the parametrization suggested by Perdew, Burke and Ernzerhof\cite{perdew96} (PBE) was used throughout. In the spirit of the DFT-D approach, the lack of vdW interactions in this semi-local functional is approximately corrected with an additional analytical, two-body inter-atomic potential. At long range, this potential equals the leading $C_{6}R^{-6}$-term of the London series, where $R$ is the inter-atomic distance, and $C_{6}$ a so-called dispersion coefficient. At short range, this long range potential is matched to the DFT inter-atomic potential by a damping function $f(R,R^{0})$, typically modulated by the vdW radii $R^{0}$ of the atom pair. In this work we use the material-specific $C_{6}$ and $R^{0}$ parameters, as well as the damping function form suggested by Tkatchenko and Scheffler \cite{tkatchenko09} (henceforth denominating corresponding results as PBE+TS). This scheme accounts to some degree for the bonding environment through a Hirshfeld-analysis based adjustment of the $C_6$ parameters, which in our previous work on azobenzene at coinage metal surface gave a superior performance compared to other DFT-D schemes \cite{mcnellis09_2,mercurio10}. 

The analytical form of the dispersion correction potential brings the advantage that dispersion-corrected total energies of geometries that are fully relaxed with respect to the dispersion-corrected forces can be obtained (employing an in-house extension to the CASTEP code) at essentially the same computational cost as a regular GGA calculation. On the other hand the assumption of a simple two-body form for the dispersion potential inherently neglects the effect of electronic screening of the vdW interactions\cite{rehr75}, which particularly for the here studied adsorption at metal surfaces is expected to lead to an overestimation of the binding energy \cite{mercurio10}. A second limitation of the semi-empirical DFT-D approach might arise for adsorbate molecules which also interact covalently with the substrate. This typically leads to molecule-substrate bond distances that are so short that the uncertainties in the heuristic damping function of the dispersion term may mingle in an uncontrolled way with deficiencies of the employed semi-local DFT xc functional.

The surface calculations were performed within supercells, using (111)-oriented metal slabs with $(6 \times 5)$ surface unit-cells and at least 18\,{\AA} vacuum. We verified that lateral interactions between adsorbed TBA and its periodic images are negligble at the GGA-PBE level. In the dispersion correction potential corresponding interactions were also switched off, so that our calculations should give a fair account of TBA adsorption in the low-coverage limit. As in our preceding work we neglected the subtle effects of the long-range Au(111) surface reconstruction. Full geometry optimizations (to within residual forces below 30 meV/{\AA}) of all molecular degrees of freedom were correspondingly performed with TBA adsorbed on one side of a four layer bulk-truncated slab. Test calculations with appropriately saturated {\em tert}-butyl groups indicated only a weakly expressed site specificity of these TBA functional groups at Au(111). This suggests that the photochemically active diazo-bridge plays a prominent role in anchoring the molecule at the metal substrate. The geometry optimizations were correspondingly initiated with the TBA diazo-bridge laterally placed as in the previously determined optimal adsorption geometry of azobenzene, corresponding to a 1:1 N-metal atom coordination \cite{mcnellis09_2}. Harmonic vibrational spectra of the adsorbed species in these relaxed geometries (and in the gas-phase) were subsequently obtained from numerical Hessians calculated by finite differences and neglecting any degrees of freedom of substrate atoms. To efficiently parallelize over the 432 displacements required for each spectrum, we have interfaced with the Atomistic Simulation Environment including the 
'phonopy' extension to it \cite{ASE_phonopy}. Not aiming to reproduce the HREELS intensities \cite{ovari07}, simple Lorentzian broadening with a width of 1 meV was applied to the spectra for better visualization.

For the energetic and electronic structure calculations the thus determined relaxed adsorbate geometries were inverted in the bottom layer, forming inversion-symmetric seven layer slabs with adsorbates at both sides. This zeroes the internal dipole moment of the slab and results in a substantially improved substrate electronic structure. The two central energetic quantities obtained with the resulting structures are the adsorption energy
\begin{equation}
E_{{\rm ads}} \;=\; \frac{1}{2}\left[E_{{\rm azo@(111)}}-E_{(111)}\right]-E_{{\rm azo(gas)}} \quad,
\label{eq:E_ads}
\end{equation}
and the relative stability of adsorbed {\em cis} (C) and {\em trans} (T) conformer
\begin{equation}
\Delta E_{\mathrm{C-T}} \;=\; \frac{1}{2} \left[E_{{\rm azo@(111)}}(\mathrm{C}) - E_{{\rm azo@(111)}}(\mathrm{T})\right] \quad.
\label{eq:delta_E}
\end{equation}
Here $E_{\rm azo@(111)}$ is the total energy of the relaxed, double-sided azobenzene-surface system, $E_{(111)}$ the total energy of the clean slab, and $E_{\rm azo(gas)}$ the total energy of the correspondingly relaxed gas-phase isomer (computed within $\Gamma$-point sampled $(35\,{\rm \AA} \times 45\,{\rm \AA} \times 35\,{\rm \AA})$ supercells). Where applicable, TS DFT-D corrections calculated in optimized four layer slab and gas-phase geometries are added to $E_{{\rm azo@(111)}}$ and $E_{\rm azo(gas)}$, respectively. Both quantities were also consistently zero-point energy corrected with the previously obtained vibrational frequencies. In the convention of Eq. (\ref{eq:E_ads}) the adsorption energy of either {\em cis} or {\em trans} isomer at the surface is thus measured relative to its stability in the gas-phase at both pure PBE and dispersion corrected PBE+TS levels of theory, and a negative sign indicates that adsorption is exothermic. Consistently, a negative sign of $\Delta E_{\mathrm{C-T}}$ indicates a higher stability of the {\em cis} isomer. Convergence tests show that at the employed plane wave cutoff of 450\,eV and $(2 \times 3 \times 1)$ Monkhorst-Pack (MP) grid\cite{monkhorst76} both energetic quantities are numerically converged to within $\pm 30$\,meV.

\subsection{Experiment}

The TPD measurements were carried out under ultrahigh vacuum conditions. The Au(111) crystal was mounted on a liquid nitrogen cooled cryostat, which in conjunction with resistive heating enables temperature control from 90\,K to 750\,K. The crystal was cleaned by a standard procedure of Ar$^{+}$ sputtering and annealing. The TBA was dosed by means of a home-built effusion cell held at 380\,K at a crystal temperature of 260\,K. In the TPD experiments, the substrate was resistively heated with a linear heating rate of 1 K/s and desorbing TBA was detected with a quadrupole mass spectrometer at the TBA-fragment mass of 190
amu (3,5-di-{\em tert}-butyl-phenyl ion). This procedure was repeated for different dosing times corresponding to different initial TBA-coverages.

As further discussed below, three desorption features ($\alpha_{1}$--$\alpha_{3}$) are observed in the TPD. They are assigned to desorption from the  multilayer ($\alpha_{1}$) and the first monolayer (ML) ($\alpha_{2}$ + $\alpha_{3}$), where $\alpha_{2}$ represents the desorption of $\approx$ 10\% of the monolayer coverage (for details see Fig.~\ref{fig5} and Ref.~\onlinecite{hagen07}). The NEXAFS and HREELS measurements of Refs. \onlinecite{schmidt10,ovari07} discussed in Section III were performed at a coverage of 1.0 and 0.9\,ML, respectively, which were prepared by heating the multilayer-covered surface to 340\,K or to 420\,K.

\section{Results and Discussion}

\subsection{Adsorption Geometry}

\begin{figure}
\begin{centering}
\includegraphics[width=1\columnwidth]{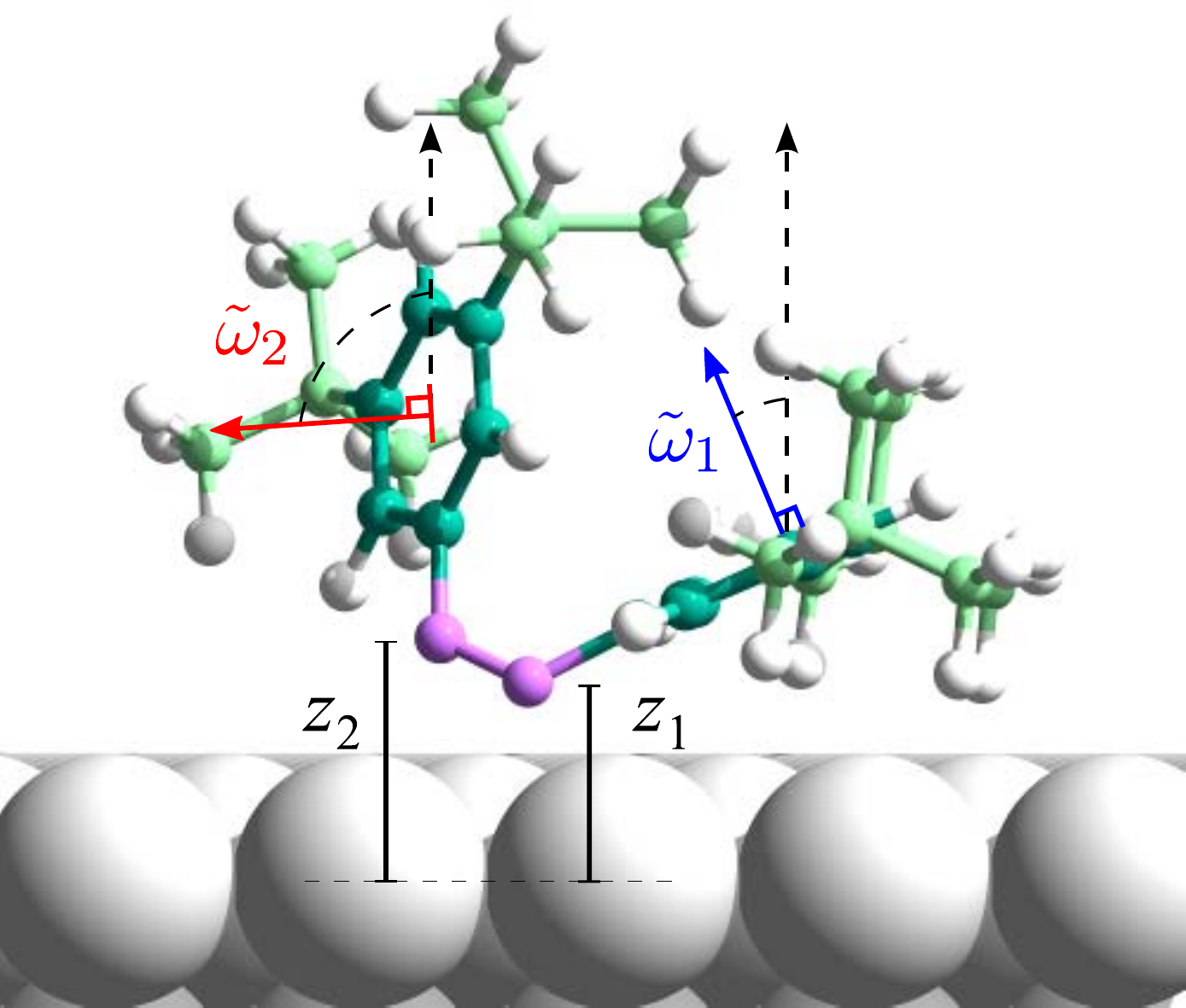}
\end{centering}
\caption{\label{fig2}
Side view of adsorbed {\em cis} TBA at Au(111), illustrating key structural parameters defining the adsorption geometry (see text): The vertical height $z_i$ of the two diazo-bridge N atoms, as well as the out-of-horizontal phenyl plane bend angles $\tilde{\omega}_i$.}
\end{figure}

Our previous work on azobenzene at coinage metal surfaces points to an understanding of the surface chemical bond in terms of a balance of four major contributions: A covalent bond between the diazo-bridge and the metal, the vdW attraction between the metal and the phenyl-rings, the Pauli repulsion between the phenyl-rings and the metal, and the energetic penalty due to the distortion of the gas-phase molecular geometry \cite{mcnellis09,mcnellis09_2}. This understanding should largely carry over to TBA at Au(111), with the bulky tert-butyl groups particularly adding to the vdW attraction and the molecular deformation upon adsorption. Key structural parameters to characterize the adsorption geometry for both {\em trans} and {\em cis} isomer are therefore the location and orientation of the central diazo-bridge, as well as the orientation of the planes of the two phenyl-rings with respect to the surface. As further illustrated in Figs. \ref{fig1} and \ref{fig2} we concentrate on the $-$N=N$-$ bond length $d_{\mathrm{NN}}$ and the vertical N atom$-$surface plane distances $z_{i}$ for the prior. Here, the indices $i=1$ and $i=2$ represent the value for the N atom connected to the phenyl-ring closer to and further away from the surface, respectively. Obviously, $z_1 = z_2$ reflects an diazo-bridge that is oriented parallel to the surface, as we obtain for the more symmetric {\em trans} isomer throughout. To specify the position of the phenyl-rings we focus on the CNNC dihedral angle $\omega$ and the out-of-horizontal bend angles $\tilde{\omega}_{i}$ of the two ohenyl planes, with the same convention for the index $i = 1,2$ to distinguish the closer and more distant phenyl-ring in the asymmetric adsorption geometry of. Fig. \ref{fig2}. Adapting to the convention employed in Refs. \onlinecite{mercurio10} and \onlinecite{schmidt10}, a dihedral angle of $\omega = 180^{\circ}$ indicates a planar TBA molecule, while $\tilde{\omega}_{i} = 0^{\circ}$ denotes a phenyl-ring that lies parallel to the surface plane. 

\begin{table}
\begin{centering}
\begin{tabular}{l|ccccc}
& \multicolumn{5}{c}{{\em Trans} @ Au(111)} \\
                & \multicolumn{2}{c}{$z_{1}=z_{2}$ ({\AA})} &  $\omega$ ($^{\circ}$) & \multicolumn{2}{c}{$\tilde{\omega}_{1}=\tilde{\omega}_{2}$ ($^{\circ}$)} \\ \hline
TBA (PBE)       & \multicolumn{2}{c}{3.97}                  & 172                    &  \multicolumn{2}{c}{9}           \\
TBA (PBE+TS)    & \multicolumn{2}{c}{3.22}          & 169                   &  \multicolumn{2}{c}{12}          \\
TBA (Exp.)\cite{schmidt10}      & \multicolumn{2}{c}{$-$}           & $-$                   &  \multicolumn{2}{c}{$4\pm5$}    \\[0.1cm]
Azo (PBE+TS)\cite{mcnellis09_2}    & \multicolumn{2}{c}{3.28}          & 180                   &  \multicolumn{2}{c}{3}           \\ \hline
& \multicolumn{5}{c}{{\em Cis} @ Au(111)} \\
             & $z_{1}$ ({\AA}) & $z_{2}$ ({\AA}) & $\omega$ ($^{\circ}$) & $\tilde{\omega}_{1}$ ($^{\circ}$) & $\tilde{\omega}_{2}$ ($^{\circ}$) \\ \hline
TBA (PBE)       & 3.25            & 3.74            & 10                    & 21        & 79          \\
TBA (PBE+TS)    & 2.34            & 2.85            & 8                     & 24        & 81          \\
TBA (Exp.)\cite{schmidt10}      & $-$             & $-$             & $-$                   & $30\pm5$ & $90\pm5$   \\[0.1cm]
Azo (PBE+TS)\cite{mcnellis09_2}    & 2.23            & 2.23            & 18                    & 32        & 68          \\ \hline
\end{tabular}
\end{centering}
\caption{\label{table1}
Structural parameters as defined in Figs. \ref{fig1} and \ref{fig2}. Values for adsorbed TBA at PBE+TS level of theory are compared to corresponding data obtained at PBE level of theory, as well as for adsorbed pure azobenzene \cite{mcnellis09_2}. Additionally shown are the out-of-horizontal phenyl plane bend angles $\tilde{\omega}_{i}$ as determined recently by NEXAFS measurements \cite{schmidt10}.}
\end{table}

\begin{figure}
\begin{centering}
\includegraphics[width=1\columnwidth]{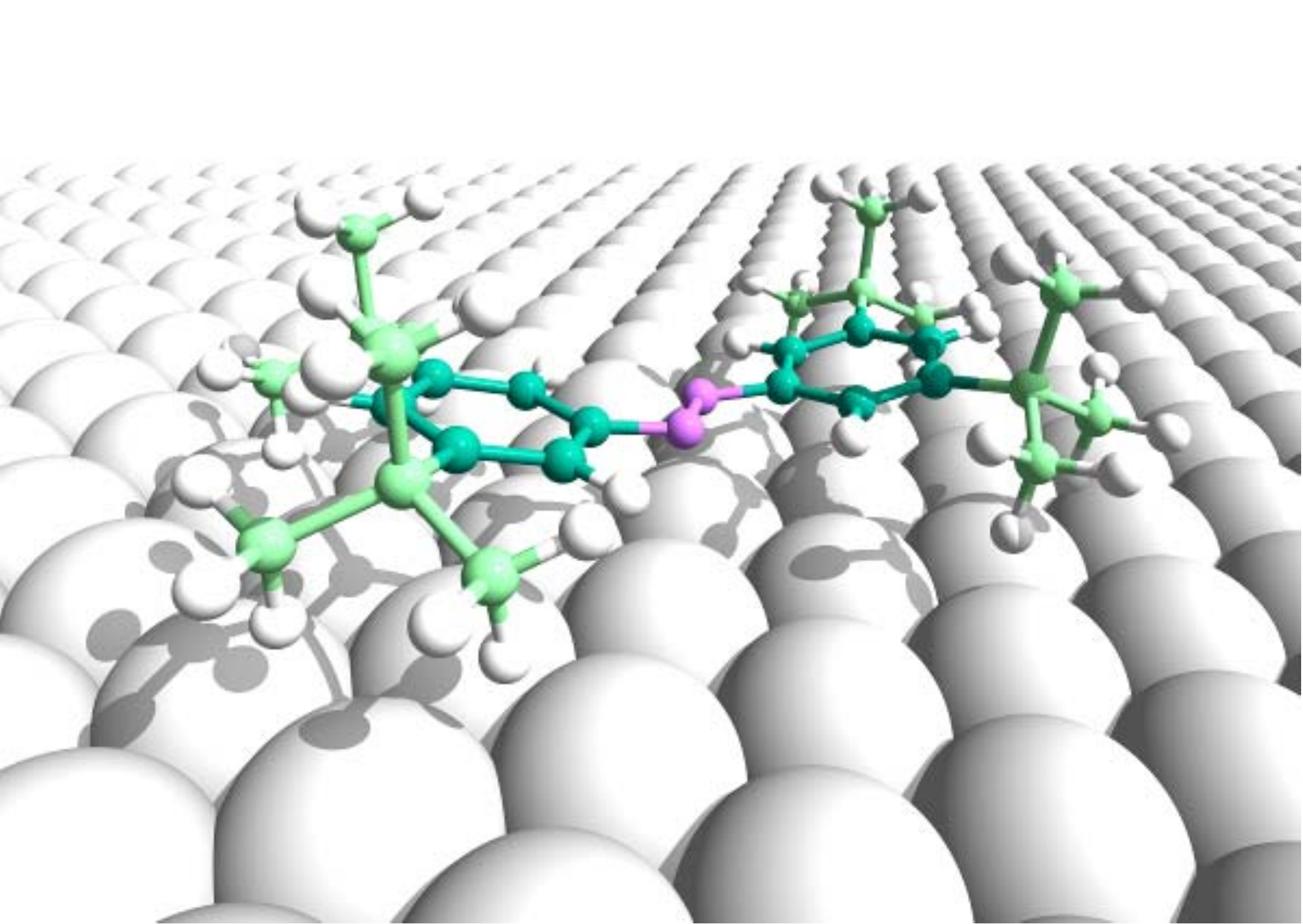}
\end{centering}
\caption{\label{fig3}
Perspective view of the {\em trans} TBA adsorption geometry at Au(111).}
\end{figure}

Table \ref{table1} summarizes the detailed geometric parameters obtained from our calculations with the exception of the diazo-bridge bond length. For the latter we consistently compute only insignificant changes away from the TBA gas-phase value of 1.29\,{\AA} ({\em trans}) and 1.28\,{\AA} ({\em cis}) at both PBE and PBE+TS level of theory. While this indicates an overall minor activation of the $-$N=N$-$ bond upon adsorption, it also demonstrates an insensivity to the degree of vdW interactions accounted for in the calculations. Starting the more detailed presentation with the {\em trans} isomer, Fig.~\ref{fig3} displays a perspective view of the overall TBA adsorption geometry. In line with the interpretation of STM \cite{alemani06,comstock07,alemani08}, surface vibrational \cite{ovari07} and NEXAFS \cite{schmidt10} data, the central azobenzene moiety essentially maintains its gas-phase planarity, with its long axis aligned to about $5^\circ$ along the direction of close-packed atom rows on the (111) surface. Also consistent with the preferential orientation deduced from NEXAFS the tert-butyl groups contact the surface with one C-H bond pointing towards the substrate. At the pure PBE level of theory, the central diazo-bridge correspondingly floats at a rather high height of about 4\,{\AA} parallel to the surface. This is further away than for pure azobenzene, where this height was 3.5\,{\AA} \cite{mcnellis09_2}. With the PBE functional not providing attractive vdW components to the molecule-surface interaction, the main effect of the bulky tert-butyl groups conforms therefore with the intuitive expectation to simply lift the functional azobenzene backbone further away from the surface. 

This picture is obviously prone to change, once some account of vdW attraction is added to the theoretical description. At the PBE+TS level of theory, the height of the diazo-bridge is indeed significantly reduced to 3.22\,{\AA}, which is nevertheless still large on the scale of a typical N-transition metal bond length. In contrast, the differential interaction with the surface is not significantly altered by the additional bonding component. The slight bending of the azobenzene moiety away from the ideal gas-phase planarity is basically the same in both the PBE and PBE+TS adsorption geometry. Here, the parallel orientation of the diazo-bridge to the surface plane and the corresponding slight upward bending of both phenyl-rings by $\sim 10^\circ$ are in very good agreement to the recent NEXAFS experiments \cite{schmidt10}. Indirectly, this corroborates some of the assumptions made in the NEXAFS data analysis, without which twice as large tilt angles would have resulted \cite{schmidt10}. On the other hand, due the insensitivity of both structural parameters due to the degree of dispersive interaction in the calculations this agreement with experiment does unfortunately not allow any conclusion as to the accuracy of the description provided by the semi-empirical TS scheme. This is particularly unfortunate, as the latter predicts an intriguing similarity of the diazo-bridge height for TBA (3.22\,{\AA}) and pure azobenzene (3.28\,{\AA}), cf. Table \ref{table1}. If correct, this obviously largely contradicts the afore mentioned intuitive view of the bulky spacer groups in terms of 'table legs' that decouple the photochemically active unit in a geometric sense.

\begin{figure}
\begin{centering}
\includegraphics[width=1\columnwidth]{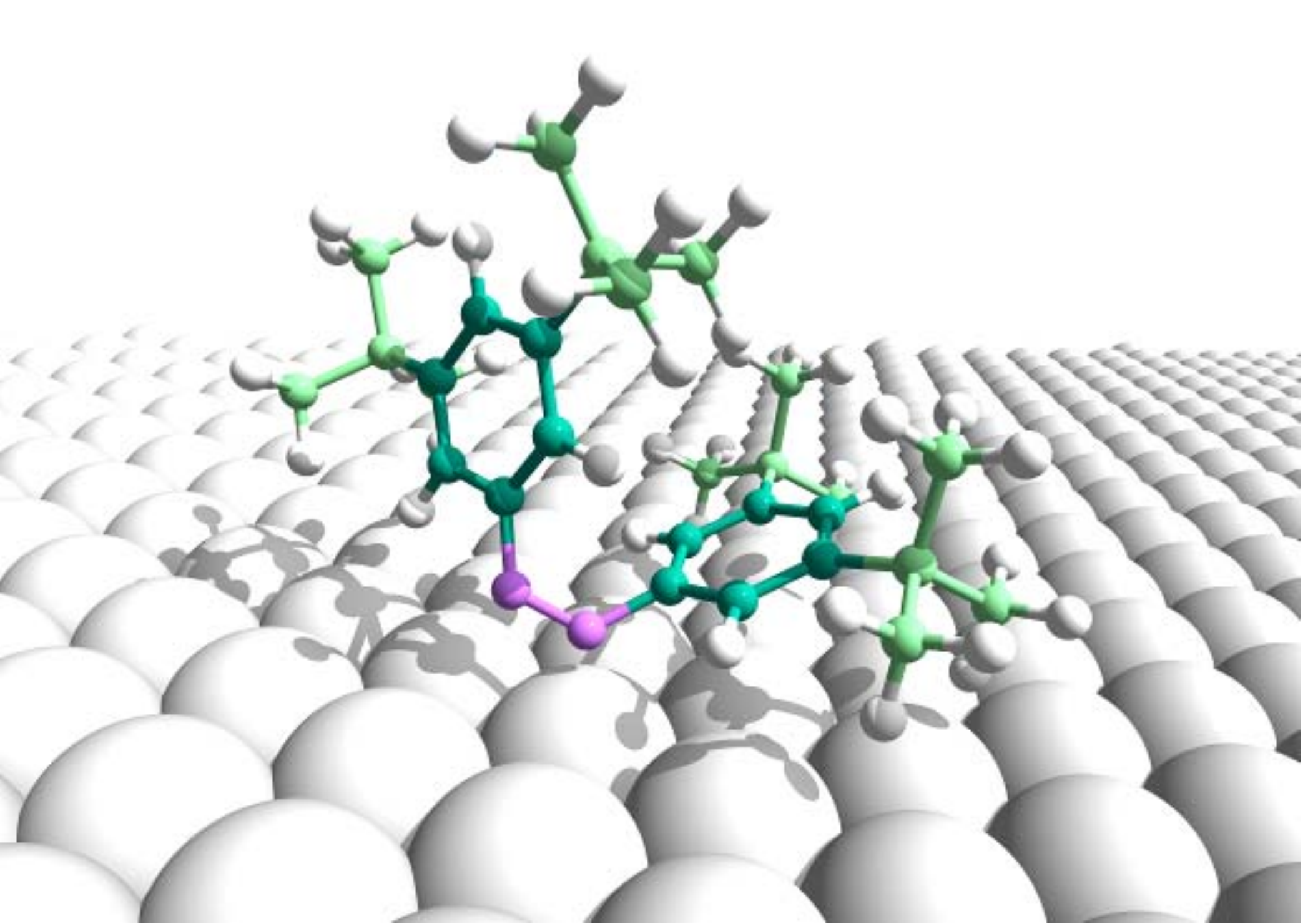}
\end{centering}
\caption{\label{fig4}
Perspective view of the {\em cis} TBA adsorption geometry at Au(111).}
\end{figure}

The similarity between functionalized and pure molecule does not extend to the adsorption geometry of the {\em cis} isomer at Au(111). As shown in Fig. \ref{fig4} this geometry is skewed for TBA with both the position and orientation of the two phenyl-rings distinctly different. This feature is consistently obtained at both PBE and PBE+TS level of theory, while for pure azobenzene both theoretical descriptions agreed on an adsorption structure with only a small sideward tilt away from a $C_2$ rotational symmetry around the central diazo-bridge. In the asymmetric adsorption mode of TBA the lower phenyl-ring is tilted out of the surface plane, yet without significant torsion that would place its two butyl groups at different heights above the surface. This is much different for the upper phenyl-ring, which does not only stand essentially upright, but is so torsioned that one of its butyl groups points largely towards the diazo-bridge and the other one away from it. Overall, the internal structure of the azobenzene backbone in this adsorbed state is therefore very similar to that in gas-phase TBA with even the CNNC dihedral angle $\omega$ not much affected by the surface potential. Another noteworthy feature of the skewed adsorption mode is the prominent tilt of the central diazo-bridge, i.e. in contrast to adsorbed {\em trans} and {\em cis} azobenzene and adsorbed {\em trans} TBA the two N atoms are at a notably different vertical height, cf. Table \ref{table1}. Qualitatively, this overall structure is perfectly consistent with the understanding reached in the recent STM \cite{alemani06,comstock07,alemani08} and NEXAFS \cite{schmidt10} measurements. As to the NEXAFS, the agreement is even quantitative in all respects. The determined out-of-surface-plane bend angles of both phenyl-rings agree very well with the assignments made, cf. Table \ref{table1}. This also extends to the inclination of the central CNNC plane with respect to the surface plane\cite{footnote}. For this, NEXAFS determines about $60^\circ$, close to the $61^\circ$ and $68^\circ$ determined at the PBE and PBE+TS level, respectively.

To some extent unfortunate and similar to the situation for the {\em trans} conformer, none of these qualitative features, as well as tilt and inclination angles are very sensitive to the description of vdW interactions in the calculations. Both PBE and PBE+TS give essentially identical results. The major feature introduced again by the account of vdW attraction is an essentially rigid downward shift of the entire molecule by about 0.9\,{\AA}. This is slightly larger than is the case for {\em trans} TBA ($\sim 0.8$\,{\AA}) and leads at least for the lower N atom of the diazo-bridge to a vertical height above the surface (2.34\,{\AA}) that is close to the value determined for pure {\em cis}-azobenzene (2.23\,{\AA}). With such prominent differences between PBE and PBE+TS, measurements of the vertical heights as was done by NI-XSW for pure azobenzene \cite{mercurio10} would therefore again provide a critical benchmark to judge on the accuracy of the account of vdW interactions introduced by the semi-empirical TS scheme. In turn, however, precisely the lacking sensitivity of the qualitative features, as well as tilt and inclination angles of both {\em trans} and {\em cis} adsorption geometries suggests that they are not much affected by the shortcoming of the employed semi-local xc functional with respect to dispersive interactions. This increases the confidence that a quite reliable understanding of the adsorption geometry of both conformers has been reached by the present calculations.

\subsection{Energetics}

\begin{figure}
\includegraphics[width=0.8\columnwidth]{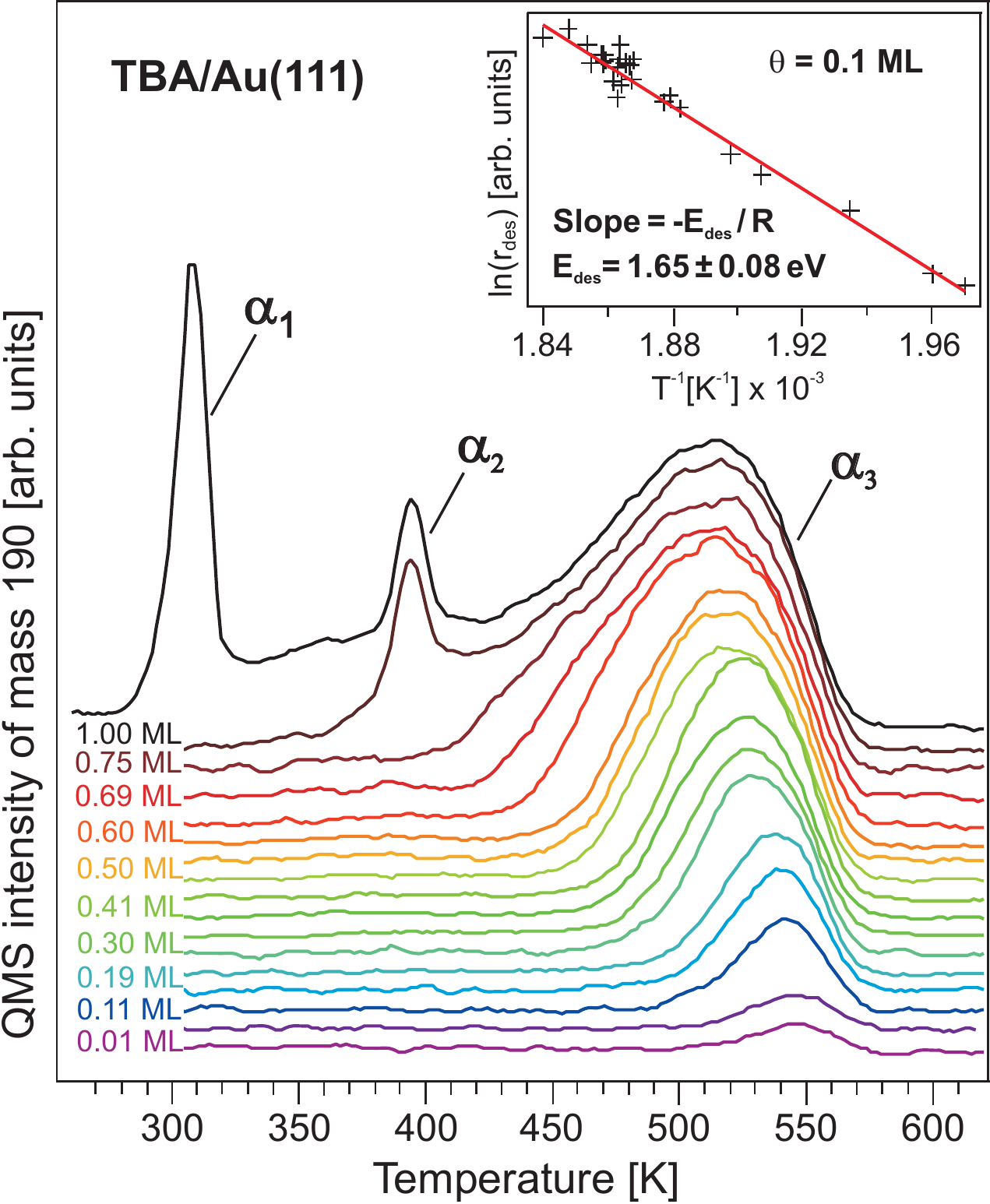}
\caption{Thermal desorption spectra (raw data) of TBA adsorbed on Au(111) at different coverages as recorded with a linear heating rate of 1 K/s at the fragment-mass of 190 amu ((C$_{4}$H$_{9})_{2}$-C$_{6}$H$_{4}^{+}$). The inset shows the desorption rate (ln$(d r_{\rm des}/dt$) plotted against the reciprocal temperature at a coverage $\theta$ of 0.1 monolayer (ML). The slope of the line, -$E_{\rm des}/R$ with $R$ the gas constant, then determines the activation energy for desorption $E_{\rm des}$ at this coverage, in this case $E_{\rm des}$(0.1 ML) = 1.65$\pm$0.08 eV.}
\label{fig5}
\end{figure}

In view of the preceding study on azobenzene \cite{mercurio10} it is clear that the binding energetics provided by the semi-empirical DFT-D approach deserves particular scrutiny. We performed TPD measurements to obtain an experimental reference value for the binding energy of the thermodynamically favored {\em trans} TBA on Au(111). Figure \ref{fig5} shows the TPD spectra as a function of TBA surface coverage recorded at the fragment mass of 190 amu ((C$_{4}$H$_{9})_{2}$-C$_{6}$H$_{4}^{+}$), with a linear heating rate of 1\,K/s. In the low coverage regime a broad desorption peak ($\alpha_{3}$) is observed around 542\,K which extends to lower temperature with increasing coverage, as has also been observed for TBA on Ag(111) \cite{tegeder07} and other azobenzene (derivatives) on noble metal surfaces \cite{ovari08,mercurio10}. This behavior is attributed to repulsive interactions between the adsorbed molecules (for example due to dipole-dipole interactions).

After saturation of the desorption peak $\alpha_{3}$ a second feature $\alpha_{2}$ develops at 394\,K. Further increase in coverage leads to saturation of this peak and to the appearance of a new feature around 308\,K labeled as $\alpha_{1}$. The $\alpha_{1}$ peak increases in height and width with increasing coverage, showing a typical zero-order desorption behavior (data not shown). We therefore assign the $\alpha_{1}$ peak to desorption from the multilayer, while $\alpha_{2}$ and $\alpha_{3}$ are associated with desorption from the monolayer. The $\alpha_{2}$ component represents the desorption of about 10 \% of the monolayer coverage and is attributed to desorption out of a densely packed TBA structure \cite{hagen07}.

In order to derive the activation energy for desorption, $E_{\rm des}$, of TBA on Au(111) in the low-coverage regime from this TPD data we utilize the so-called complete analysis \cite{christmann91}, which has the advantage that knowledge about the absolute coverage is not required. For this analysis a family of TPD curves are measured as a function of initial coverage ($\theta_{i}$) as shown in Fig. \ref{fig5}. These curves are subsequently used to construct a family of $\theta(t)$ curves via $A_{p}\propto \int_{0}^{\infty} r_{\rm des}(t)dt= \theta$ with $A_{p}$ the peak area and $r_{\rm des}$ the desorption rate. Due to the known linear heating rate (in our case 1\,K/s), this corresponds to a knowledge of $\theta(T_{s}$) as a function of $\theta_{i}$, where $T_{s}$ is the surface temperature. Following this, an arbitrary coverage value $\theta_{1}$ is chosen that is contained in each of the desorption curves. The desorption rate at this coverage, $r_{\rm des}(\theta_{1})$, and the temperature at which this rate was obtained, $T_{1}$, are then read off from each of these desorption curves. Plotting ln$[r_{\rm des}(\theta_{1}$)] versus $1/T_{1}$ (Arrhenius-plot) finally allows to determine $E_{\rm des}(\theta_{1}$) as follows directly from the Polanyi-Wigner equation \cite{christmann91}. This is exemplarily shown for a coverage of 0.1\,ML in the inset of Fig. \ref{fig5}, which yields an activation energy of 1.65$\pm$0.08\,eV. 

\begin{table}
\begin{centering}
\begin{tabular}{ll|ccc}
          &        & {\em Trans}        &                           & {\em Cis}          \\
          &        & $E_{\mathrm{ads}}$ & $\Delta E_{\mathrm{C-T}}$ & $E_{\mathrm{ads}}$ \\ \hline
Gas-phase & PBE    & $-$                &   0.58                    &   $-$              \\
          & PBE+TS & $-$                &   0.29                    &   $-$              \\ \hline
@ Au(111) & PBE    & $-0.16$            &   0.52                    & $-0.21$            \\
          & PBE+TS & $-3.00$            &   0.99                    & $-2.07$            \\ \hline
Experiment&        & $-1.70\pm0.1$      &   $-$                     & $-$                \\ \hline
\end{tabular}
\par\end{centering}
\caption{\label{table2}
Energetics of TBA in gas-phase, and adsorbed at Au(111). All numbers are in eV.}
\end{table}

In the low-coverage regime $\leq$ 0.2 ML this analysis determines an essentially constant desorption energy of $E_{\rm des}=1.70\pm$0.1\,eV. Assuming no additional activation barrier for adsorption, this then provides the aspired benchmark number against which the calculated $E_{\rm ads}$ for {\em trans} TBA may be referenced. As apparent from Table \ref{table2} the values obtained at the PBE and PBE+TS level are very consistent with the findings of the previous studies of benzene and pure azobenzene at coinage metal surfaces \cite{mcnellis09_2,mercurio10}: The lack of vdW interactions in the semi-local PBE functional leads to an overall negligible binding of both TBA isomers, with the actual values for $E_{\rm ads}$ in fact essentially identical to those computed for pure azobenzene at Au(111) before \cite{mcnellis09_2}. Adding dispersive attraction within the DFT-D approach these binding energies are dramatically increased, such that in the end the intended dispersion 'correction' amounts to more than 90\,\% of the total binding energy. Compared to the experimental reference the PBE+TS approach overbinds the {\em trans} isomer by about as much as PBE underbinds. In absolute numbers this is a disconcerting deviation of more than 1\,eV. In our previous work we had assigned the corresponding overbinding determined for pure azobenzene to the neglect of metallic screening in the DFT-D approach\cite{mercurio10}. This argument was largely based on the intriguing accuracy of the PBE+TS adsorption geometry compared to the bond distances derived from NI-XSW measurements. As discussed above, such a conclusion on the determined TBA adsorption geometry is presently not possible, as all hitherto measured structural parameters are not very sensitive to the additional vdW attraction provided by the PBE+TS scheme. On the other hand, the additional butyl groups lead to a significantly increased overall binding energy of TBA compared to azobenzene, and in both cases PBE+TS overbinds by roughly 40\,\% (azo: 1.71\,eV vs. 1.08\,eV \cite{mercurio10}; TBA: 3.00\,eV vs. 1.70\,eV). Such a rather geometry-unspecific deviation between theoretical and experimental data quite well fits the anticipated effect of an overestimated uniform background potential. The latter is attributed to the neglect of screening of vdW contributions from more distant substrate atoms. 

For gas-phase TBA PBE+TS yields a somewhat smaller {\em cis}-{\em trans} energy difference $\Delta E_{\rm C-T}$ than for pure azobenzene, 0.29\,eV compared to 0.49\,eV\cite{mcnellis09_2}, respectively. This arises predominantly from the additional vdW attractions due to the butyl groups in the bend {\em cis} conformer, cf. Fig. \ref{fig1}, and is not found at the PBE level of theory, where $\Delta E_{\mathrm{C-T}}=0.58$\,eV for both TBA and azobenzene. Compared to the gas-phase reference, the stability difference of the two TBA conformers is with 0.99\,eV substantially increased at Au(111) at PBE+TS level of theory. This is primarily due to the larger vdW attraction possible in the planar {\em trans} adsorption mode and was equivalently obtained for pure azobenzene adsorption \cite{mcnellis09_2}. This interpretation in terms of vdW is corroborated by the essentially unchanged $\Delta E_{\mathrm{C-T}}$ of gas-phase and adsorbed TBA at PBE level of theory, where these bonding contributions are absent, cf. Table \ref{table2}. Not withstanding, in light of the suspected overestimation of the vdW attraction within PBE+TS we also expect that this increase of $\Delta E_{\mathrm{C-T}}$ upon adsorption is overestimated. Extrapolating the roughly 40\,\% overshoot seen in the {\em trans} TBA binding energy, we would therefore cautiously conclude on only a moderate $\sim 0.3-0.4$\,eV increase of the {\em cis}-{\em trans} energy difference upon adsorption of TBA at Au(111). While the therewith implied higher stability of adsorbed {\em trans} TBA is consistent with the existing experimental data, there are to date unfortunately no dedicated measurements of $\Delta E_{\mathrm{C-T}}$ against which this estimate could be compared.

\subsection{Vibrations}

\begin{figure}
\includegraphics[width=1.0\columnwidth]{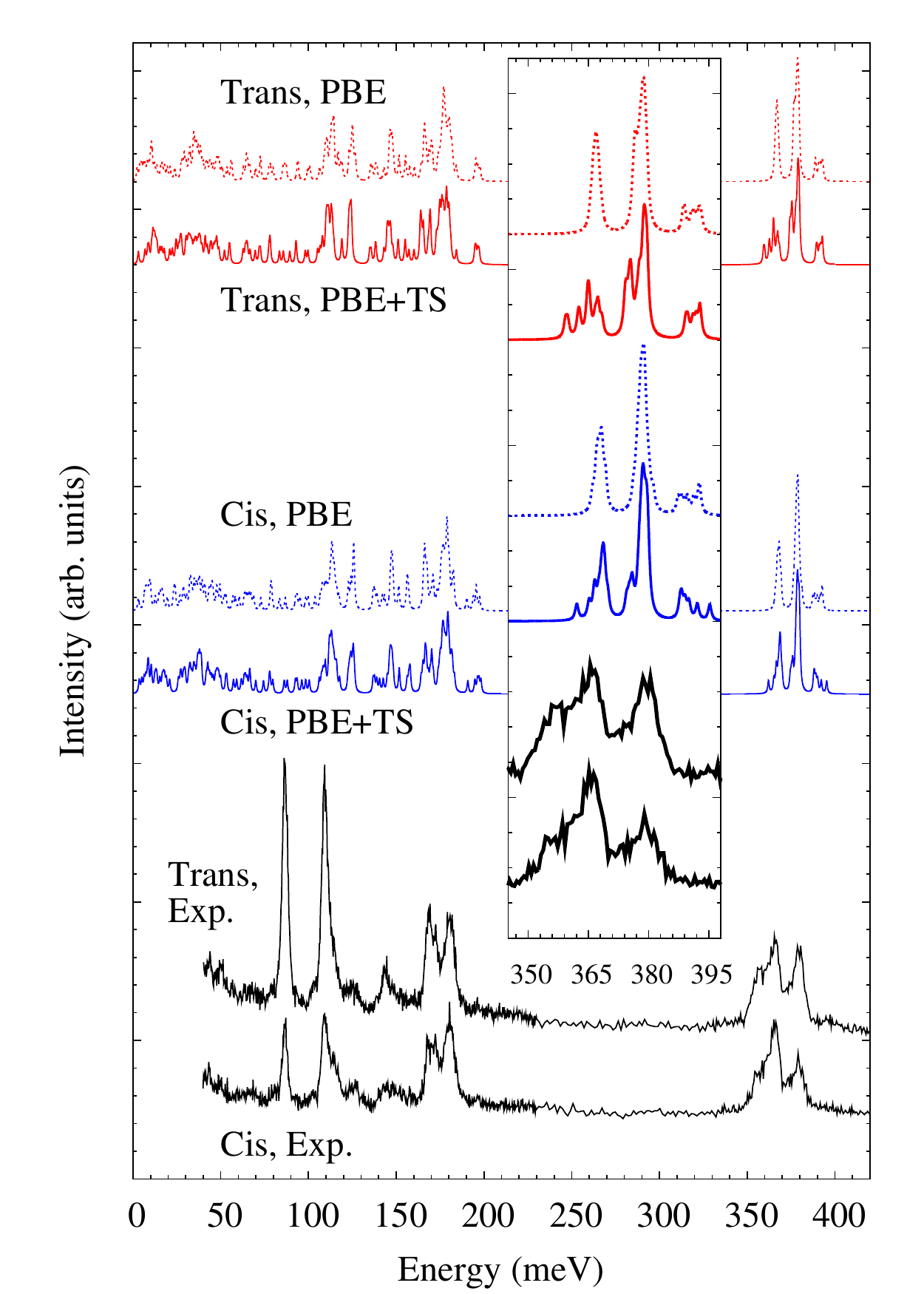}
\caption{Comparison of calculated surface vibrational modes of {\em trans} (top, red lines) and {\em cis} (center, blue lines) TBA at Au(111) at PBE (dashed lines) and PBE+TS (solid lines) level of theory to the corresponding HREELS spectra from ref. \onlinecite{ovari07} (bottom, black solid lines. Upper: {\em trans}, Lower: {\em cis}). Center inset: The highest energy peaks of the spectrum, shown at larger scale. Note the C-H stretch peak on the left. The spectra have been vertically displaced for clarity. Not aiming to reproduce the HREELS intensities, the computed modes are convoluted with a Lorentzian broadening of 1 (meV).}
\label{fig6}
\end{figure}

The picture arising from the geometric and energetic characterization of TBA at Au(111) points to a surface chemical bond that is predominantly due to unspecific dispersive interactions. Particularly for the {\em trans} conformer, the structural similarity of the azobenzene and TBA adsorption complex with respect to the photochemically active diazo-bridge moiety (together with the in both cases negligible $-$N=N$-$ bond activation) questions the oft-quoted function of the tert-butyl 'spacers' in terms of a geometric decoupling from the surface. Complementary information confirming this view can come from an analysis of the surface vibrational modes. Figure \ref{fig6} compares corresponding results computed for {\em trans} TBA at Au(111) at PBE+TS level of theory with experimental data from HREELS \cite{ovari07}. At first glance, we find overall agreement with all major features nicely reproduced by the calculations. As had already been noticed in the experimental work, the vibrational spectrum exhibits only minor changes between adsorbed {\em cis} and {\em trans} TBA and even TBA in the condensed phase \cite{ovari07,kuebler60, kellerer71}. This is similarly obtained in the calculations, as illustrated by the spectrum for {\em cis} TBA at Au(111) also shown in Fig. \ref{fig6}.

The detailed inspection of the C-H vibrational modes in the energy range of 350 to 400~meV reveals nevertheless the subtle influence of the Au(111) substrate on the vibrational spectra of the adsorbed species. Based on the calculations the three frequency bands centered at 368, 378, and 390\,meV are assigned to the symmetric and asymmetric C-H vibrations of the tert-butyl groups, as well as the C-H bending modes at the phenyl-rings, respectively. There is a clear change of the former two vibrational bands when going from the PBE to the PBE+TS description, cf. the inset of Fig. \ref{fig6}. The overall reduced distance of the TBA molecule to the gold substrate upon inclusion of the attractive vdW forces leads to a clear splitting of the C-H vibrations of the {\em tert}-butyl moieties. In the adsorbed species the frequencies of the C-H bonds pointing towards the substrate are red shifted by 5-10\,meV and the corresponding vibrations develop low-energy sidebands. These subtle but important changes are confirmed by the experimental HREELS spectra. Due to the mainly in-plane character of the C-H phenyl-ring bending modes we observe only a small shoulder at 390\,meV. In contrast the symmetric and asymmetric C-H stretch vibrations of the {\em tert}-butyl legs at 368 and 380\,meV result in pronounced dipole-active bands.  Upon {\em trans} to {\em cis} isomerization the number of C-H bonds pointing towards and noticeably contacting the substrate is roughly halved, cf. Fig. \ref{fig2}. This leads in both experiment and theory to a reduction of the low-energy side bands best seen for the symmetric C-H stretch vibrations which constitute the lowest band at 368\,meV. As the splitting of the C-H stretch vibrations happens only at a reduced distance to the substrate (PBE vs. PBE+TS adsorption structure) it confirms the importance of vdW interaction in the bonding of TBA. Unfortunately this is not a quantitative benchmark for the accuracy of our semi-empirical PBE+TS approach. 

More insight into the specific bonding of the diazo-bridge moiety can specifically come from a detailed analysis of the $-$N=N$-$ stretch mode. However, identically obtained with PBE and PBE+TS, the computed value of 184\,meV in adsorbed {\em trans} and {\em cis} TBA again reflects a negligible activation of the NN bond at the surface. Compared to the respective gas-phase values, the mode red-shifts only by about 3\,meV for both isomers. Correspondingly, the obtained shift between this stretch mode in {\em trans} and {\em cis} TBA at Au(111) is very similar to the one computed for free TBA, 6 vs. 5 meV, respectively. This in turn compares very well to the 8\,meV measured for TBA in the condensed phase with IR and Raman spectroscopy \cite{kuebler60,kellerer71}. Such a negligible change of the {\em cis}-{\em trans} stretch frequency difference upon adsorption is not found for pure azobenzene at Au(111). Here, the closer encounter of the diazo-bridge in the {\em cis} adsorption geometry softens the stretch mode by more than 6 meV. With the mode unaffected in adsorbed {\em trans} azobenzene, this leads to a concomitant increase of the {\em cis}-{\em trans} stretch frequency shift compared to the gas-phase. In {\em cis} TBA at Au(111) a corresponding softening does not occur as the diazo-bridge does not come as close to the surface in the skewed adsorption mode. In this respect, one can really attest some geometric decoupling effect to the butyl groups, yet only for the {\em cis} isomer. For the {\em trans} isomer though, an equivalent analysis of the diazo-bridge against surface stretch or other torsional azo modes shows always the same similarity between TBA and azobenzene at Au(111) as for the just discussed $-$N=N$-$ stretch mode. Also the characterization of the surface vibrational properties supports therefore the understanding that the effect of the bulky {\em tert}-butyl groups for the switching properties of {\em trans} TBA at Au(111) must be more subtle than a mere geometric decoupling of the central photochemically active molecular moiety.

\section{Summary and Conclusions}

A methodological motivation of this work was to further explore the capabilities of the semi-empirical dispersion correction approach to semi-local DFT in describing the adsorption of complex organic molecules at metal surfaces. For this we have presented a detailed characterization of the geometric, energetic and vibrational properties of the {\em trans} and {\em cis} isomers of the azobenzene derivate TBA at Au(111). The findings are in all respects in line with the experience from preceding studies on benzene and pure azobenzene at coinage metal surfaces \cite{mcnellis09,mcnellis09_2,mercurio10}: The additional account of attractive vdW interaction introduced by the PBE+TS scheme leads to a significant modification of the adsorption geometry, primarily in terms of bringing the molecule closer to the surface. The concomitantly dramatically increased binding energy is notably overestimated compared to the reference value from TPD measurements. For azobenzene at Ag(111) a corresponding overbinding -- in fact in relative terms very much comparable to the one found here for TBA at Au(111) -- was attributed to the neglect of metallic screening of dispersive interactions in the semi-empirical DFT-D approach \cite{schmidt10}. This argument was largely based on the very accurate PBE+TS adsorption geometry as compared to detailed structural data from NI-XSW measurements. Such a conclusion is, unfortunately, not yet possible in full for the here studied TBA at Au(111). The qualitative features of the determined {\em trans} and {\em cis} adsorption geometries and even detailed tilt and inclination angles agree all very well with existing data from STM \cite{alemani06,comstock07,alemani08}, HREELS \cite{ovari07} and NEXAFS \cite{schmidt10} experiments. The energetic lowering of the C-H stretch vibrations of the {\em tert}-butyl groups pointing towards the substrate is only observed in the PBE+TS approach. This points towards the importance of the vdW component introduced by the DFT-D approach. However, no experimental reference data exist to date for the vertical height of the adsorbed molecule, that would allow to directly confirm the present working hypothesis, namely that the PBE+TS approach is a very useful and computationally tractable tool to provide accurate structural information for adsorbed complex organic molecules.

If this working hypothesis proves correct and the determined PBE+TS adsorption geometries are indeed accurate, the prevailing preconception of the role of the bulky functional groups for the experimentally observed improved isomerization ability of adsorbed TBA needs revision. Motivated by the gas-phase structure, this view discusses the butyl groups as spacers that help to decouple the molecular switch from the metal substrate. For the TBA {\em cis} isomer we indeed compute a skewed adsorption mode, in which one phenyl-moiety is much more tilted, up to the point of standing essentially perpendicular to the surface plane. This is quite different to the more symmetric adsorption mode of {\em cis} azobenzene at Au(111), such that here one can attest some geometric decoupling from the surface due to the butyl 'table legs'. Quite distinctly, the almost planar adsorption mode of {\em trans} TBA is very much comparable to the adsorption geometry of pure azobenzene. Particularly for the photochemically active diazo-bridge moiety this similarity goes even down to essentially unchanged structural and vibrational properties. In this respect, a core message arising from the present study is that especially for the photo-switching of {\em trans} TBA the effect of the bulky spacer groups must be more subtle than the anticipated mere geometric decoupling of the functional backbone from the metal surface.

\section{Acknowledgements}

Funding by the Deutsche Forschungsgemeinschaft through Sfb 658 - Elementary Processes in Molecular Switches at Surfaces - is gratefully acknowledged. Ample computing time has been provided at the HLRB-II architecture of the Leibniz Supercomputing Centre in Garching.


\begin{thebibliography}{99}
\bibitem{footnote}
The central CNNC atoms of the azobenzene backbone are only precisely located within one plane for a dihedral angle $\omega = 0$. For the small, but finite angles $\omega$ determined in the calculations, we therefore use a simple arithmetic average to determine the inclination of the CNNC plane in the comparison to the NEXAFS data.
\end{thebibliography}
\end{document}